\documentclass[slac_one]{revtex4}
\usepackage{graphicx}
\usepackage{fancyhdr}
\begin{document}

\title{KLOE results on light meson properties} 

%

\author{C.Bini}

\author{the KLOE collaboration\footnote{
F.~Ambrosino,
A.~Antonelli,
M.~Antonelli,
F.~Archilli, 
P.~Beltrame,
G.~Bencivenni,
S.~Bertolucci,
C.~Bini,
C.~Bloise,
S.~Bocchetta,
F.~Bossi,
P.~Branchini,
P.~Campana,
G.~Capon,
T.~Capussela,
F.~Ceradini,
P.~Ciambrone,
F.~Crucianelli,
E.~De~Lucia,
A.~De~Santis,
P.~De~Simone,
G.~De~Zorzi,
A.~Denig,
A.~Di~Domenico,
C.~Di~Donato,
B.~Di~Micco,
M.~Dreucci,
G.~Felici,
M.~L.~Ferrer,
S.~Fiore,
P.~Franzini,
C.~Gatti,
P.~Gauzzi,
S.~Giovannella,
E.~Graziani,
W.~Kluge,
G.~Lanfranchi,
J.~Lee-Franzini,
D.~Leone,
M.~Martini,
P.~Massarotti,
S.~Meola,
S.~Miscetti,
M.~Moulson,
S.~M\"uller,
F.~Murtas,
M.~Napolitano,
F.~Nguyen,
M.~Palutan,
E.~Pasqualucci,
A.~Passeri,
V.~Patera,
F.~Perfetto,
P.~Santangelo,
B.~Sciascia,
A.~Sciubba,
A.~Sibidanov,
T.~Spadaro,
M.~Testa,
L.~Tortora,
P.~Valente,
G.~Venanzoni,
R.Versaci,
G.~Xu
}} 

\affiliation{Universit\'a ``La Sapienza'' and INFN, Roma}
%

\begin{abstract}
A review of the recent results obtained by the KLOE experiment at DAFNE 
concerning the physics of low mass mesons is presented.
\end{abstract}

\maketitle


\thispagestyle{fancy}


\section{OVERVIEW OF THE KLOE EXPERIMENT AT DAFNE} 
The KLOE experiment has been working at the $e^+e^-$
collider DAFNE, the $\phi$-factory of the Frascati Laboratories, running 
at a centre of mass energy around 1.02 GeV 
(the $\phi$ resonance peak) with a luminosity 
peak value of $1.5\times 10^{32}$cm$^{-2}$s$^{-1}$.
KLOE has collected about 2.7
fb$^{-1}$ total integrated luminosity mostly at the
$\phi$ peak (2.5 fb$^{-1}$
corresponding to about $8\times 10^9$ $\phi$ decays). The other data have been
taken around the $\phi$ between 1.00 and 1.03 GeV.

The main mission of the $\phi$-factory is the study of kaon
physics, 83\% of $\phi$
decays being in
kaon pairs. However, the $\phi$-factory is a
copious source of mesons with mass below 1 GeV, especially through 
radiative decays.

In the following, results are presented concerning the physics of the
lowest
mass scalar mesons (the f$_0$(980), the a$_0$(980) and the f$_0$(600)) of the
$\eta$-$\eta'$ mesons and of vector
mesons. An update is given also of the measurement of the
$e^+e^-\rightarrow\pi^+\pi^-$ cross-section for centre of mass energy below 1
GeV done by KLOE using the radiative return method.
Finally the prospects of KLOE and DAFNE are briefly discussed.
\section{RESULTS}

\subsection{Scalar Mesons}
The lowest mass scalar mesons are accessible at DAFNE through
$\phi\rightarrow\pi\pi\gamma$ (sensitive to the isospin 0 scalar mesons
f$_0$(980) and f$_0$(600), the ``controversial'' $\sigma$ meson) 
and $\phi\rightarrow\eta\pi\gamma$ final states (sensitive to the isospin 1
a$_0$(980)). On the other hand, both isospin 0 and 1
mesons contribute to $\phi\rightarrow K\overline{K}\gamma$. 

The study of the aforementioned decays essentially allows to measure the
coupling of the scalar mesons to the $\phi$ meson, 
to kaon pairs and to $\pi\pi$ or
$\eta\pi$. These couplings are strictly 
related to the quark composition of the mesons. Moreover the study 
of the $\pi\pi\gamma$ 
decay dynamics allows to look for effects due to the $\sigma$ meson.

KLOE has published results concerning $\pi^0\pi^0\gamma$~\cite{stesim1,stesim2}, 
$\pi^+\pi^-\gamma$~\cite{mio} and
$\eta\pi^0\gamma$~\cite{mioold}. Here we present the final results of a new analysis
of $\eta\pi^0\gamma$ corresponding to a statistics 20 times larger than
the previous one~\cite{mioold}, and the preliminary result of a direct search
of $K^0\overline{K^0}\gamma$.   

\begin{figure*}[t]
\centering
\includegraphics[width=70mm]{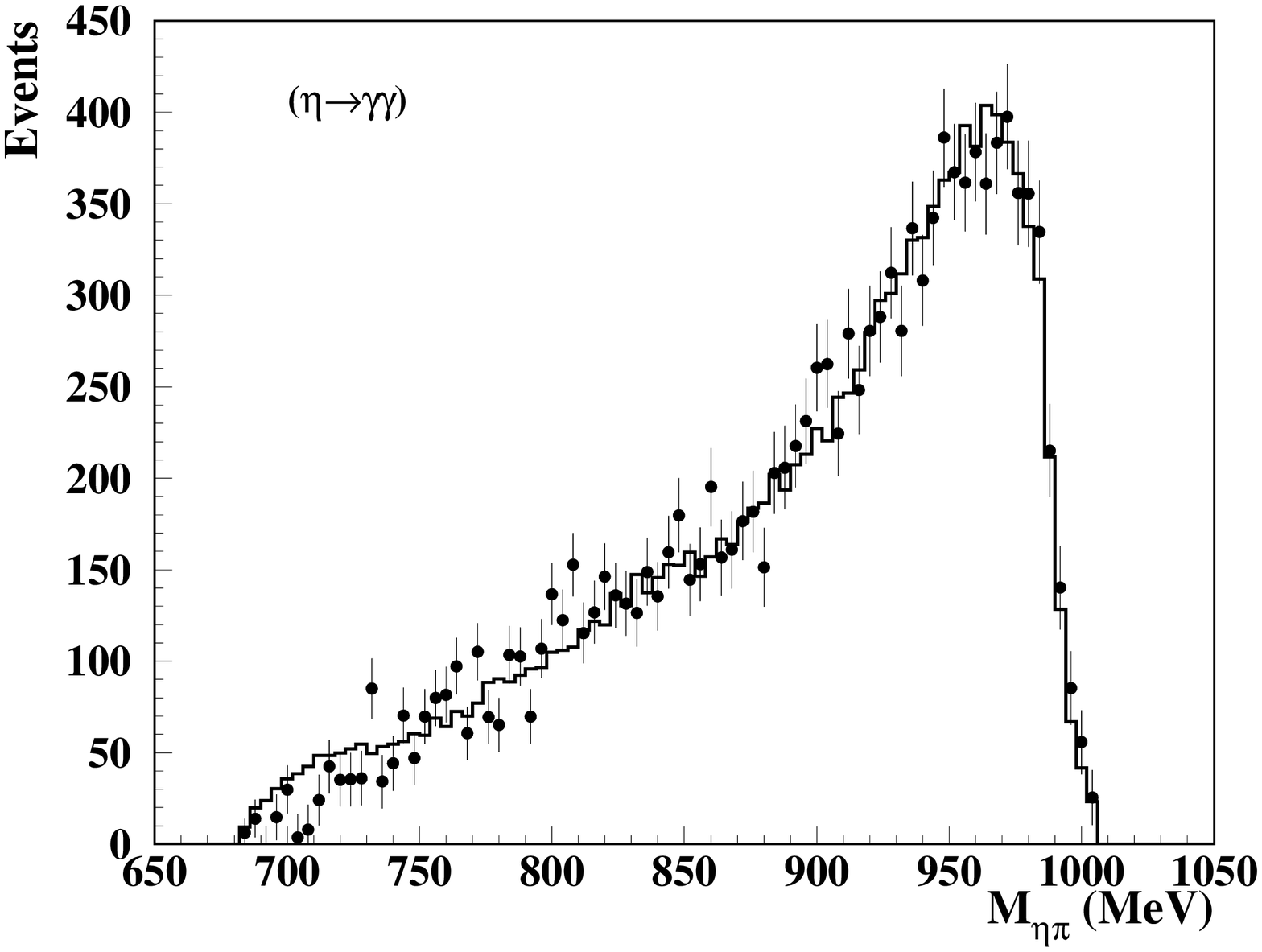}
\includegraphics[width=70mm]{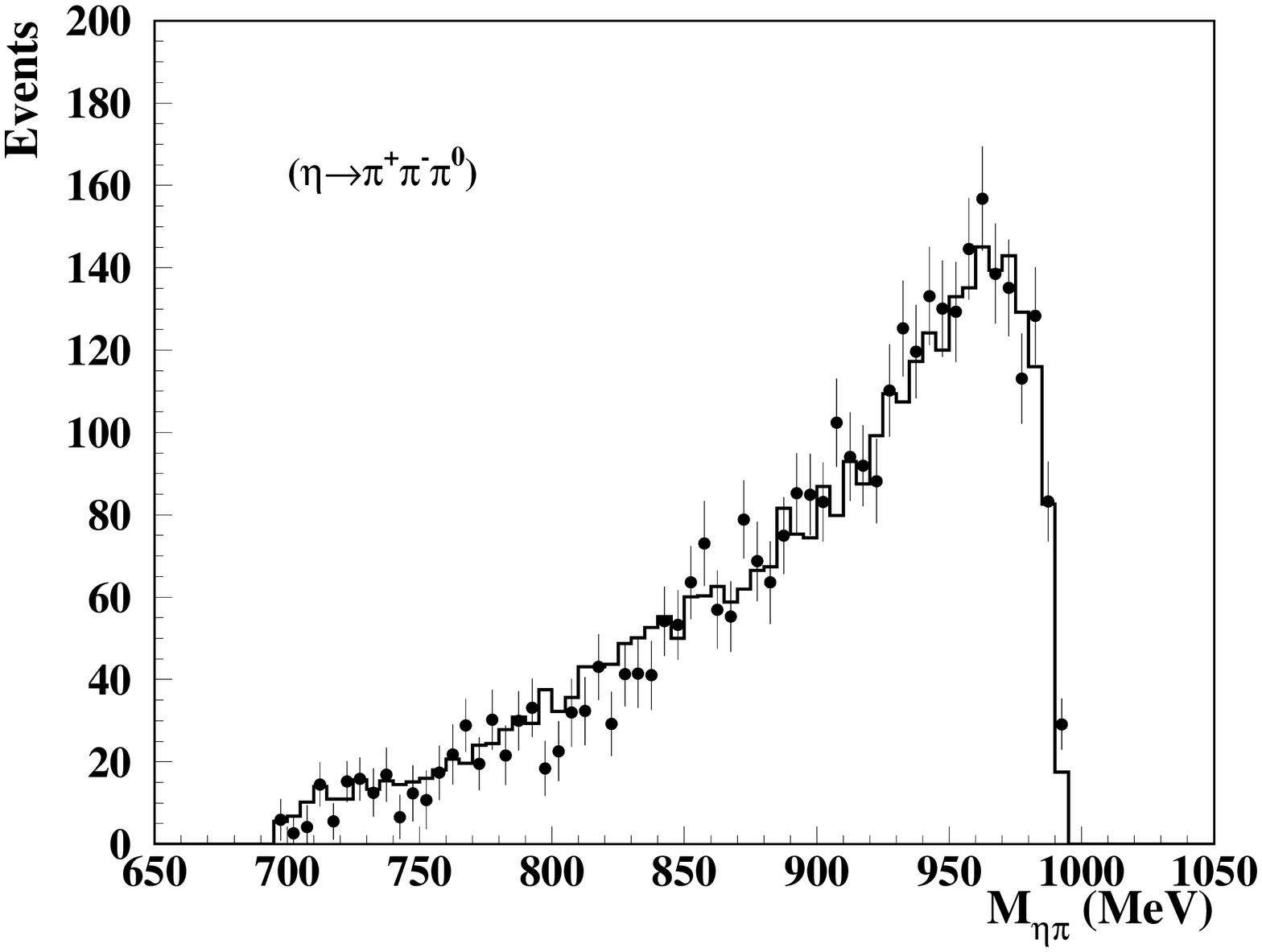}
\caption{$\eta\pi$ invariant mass spectrum of $\phi\rightarrow\eta\pi\gamma$
  for the $\eta\rightarrow\gamma\gamma$
  sample (left plot) and for the $\eta\rightarrow\pi^+\pi^-\pi^0$ sample (right
  plot). Points are data and histogram is the result of the combined fit
  with the theoretical spectrum convoluted with experimental 
  efficiency and resolution} \label{scalars}
\end{figure*}

Two samples of $\eta\pi^0\gamma$ events are selected from a data set of about
450 pb$^{-1}$: one with
$\eta\rightarrow\gamma\gamma$ and one with
$\eta\rightarrow\pi^+\pi^-\pi^0$. The first sample is fully neutral (5 photons
final state) and is characterised by large statistics (about 20000 events are selected)
but also by large reducible background mostly coming from $\phi\rightarrow\eta\gamma$
with $\eta\rightarrow3\pi^0$. The second sample  (about 4000 events
selected) gives rise
to a final state of 2 tracks and 5 photons with lower background. After
background subtraction we combine the two samples to get
the measurement of the branching ratio:
\begin{equation}
B.R.(\phi\rightarrow\eta\pi^0\gamma)=(7.05\pm0.08_{stat}\pm0.21_{syst})\times 10^{-5}
\end{equation}
A combined fit of the two $\eta\pi$ invariant mass spectra is done to
extract the shape and the parameters of the $\phi\rightarrow$a$_0\gamma$ decay
amplitude that dominates the $\phi\rightarrow\eta\pi\gamma$ decay. 
The fit (see Fig.\ref{scalars}) is done using
the kaon loop model~\cite{Achasov} to describe the scalar amplitude. 
The values of the couplings, $g_{a_0\eta\pi}$ and
$g_{a_0KK}$ can be compared to the predictions of several models~\cite{VARI}. In
particular they fit quite well a recently proposed model based on instanton
interactions~\cite{MAIANI}.  

To search for the rare $\phi\rightarrow K^0\overline{K^0}\gamma$ decay we look
for events $K_SK_S\gamma$ where both $K_S$ promptly decay to $\pi^+\pi^-$. The
background is strongly reduced by the request of two kaon vertexes with 
missing momentum matching a
low energy photon in the calorimeter. Out of a 1.4 fb$^{-1}$ sample, we find
1 event in the data and no events in a Montecarlo sample of the same size
that includes all possible backgrounds. From these data we set a
90\% C.L.
upper limit:
\begin{equation}
B.R.(\phi\rightarrow K^0\overline{K^0}\gamma)<1.8\times 10^{-8}
\end{equation}
that can be also compared to the predictions of 
several models concerning the structure of the scalar 
mesons~\cite{VARI}.

\subsection{$\eta$-$\eta'$ physics}
$\eta$ and $\eta'$ mesons are copiously produced through the radiative decays
$\phi\rightarrow\eta\gamma$ (B.R.=1.2\%) and $\phi\rightarrow\eta\gamma$
(B.R.=6.2 $\times 10^{-5}$). In both cases the monochromatic 
radiated photon allows to tag the event giving
almost background-free $\eta$ samples and good
$\eta'$ samples.

Among the measurements carried out in this field we mention the precision
measurement of the $\eta$ mass~\cite{etamass}, of the $\eta$-$\eta'$
mixing~\cite{mixing}, of the dynamics of the 3 pion $\eta$
decays~\cite{trepioni} and the observation and measurement of the branching
ratios of the rare decays $\eta\rightarrow\pi^+\pi^-e^+e^-$, $\eta\rightarrow
e^+e^-e^+e^-$ and $\eta\rightarrow\pi^0\gamma\gamma$~\cite{etarare}.

\subsection{Measurement of $\phi\rightarrow\omega\pi^0$}
\begin{figure*}[t]
\centering
\includegraphics[width=85mm]{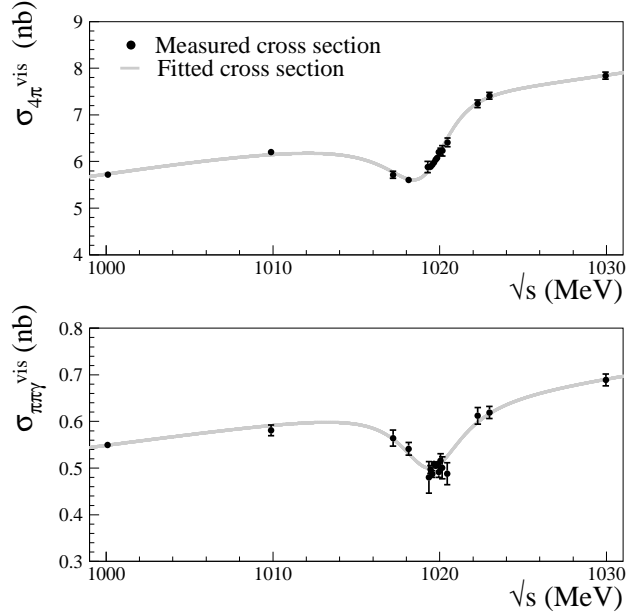}
\caption{Visible cross-section as a function of the centre of mass energy
  $\sqrt{s}$ for the processes $e^+e^-\rightarrow\pi^+\pi^-\pi^0\pi^0$ (upper
  plot) and $e^+e^-\rightarrow\pi^0\pi^0\gamma$ (lower plot) both dominated by
  the 
  $\omega\pi^0$ intermediate state. The curve is the
  result of the fit.} \label{omegapi0}
\end{figure*}

The decay $\phi\rightarrow\omega\pi^0$ is a OZI and G-parity violating
process. To detect it at KLOE, we measure the
cross-section of the processes $e^+e^-\rightarrow\pi^+\pi^-\pi^0\pi^0$ and 
$e^+e^-\rightarrow\pi^0\pi^0\gamma$, both dominated by the $\omega\pi^0$
intermediate state as a function of
the centre of mass energy $\sqrt{s}$ (see Fig.\ref{omegapi0}). 
For this analysis~\cite{omegapi0} we use the energy
scan around the $\phi$ peak between 1000 and 1030 MeV. 
From the fit shown in Fig.\ref{omegapi0} we extract the quantities:
\begin{equation}
B.R.(\phi\rightarrow\omega\pi^0)=(4.4\pm0.6)\times 10^{-5}
\end{equation}
\begin{equation}
\Gamma(\omega\rightarrow\pi^0\gamma)/\Gamma(\omega\rightarrow\pi^+\pi^-\pi^0)=0.0897\pm0.0016
\end{equation} 
Using unitarity and the PDG values~\cite{PDG} for the rare $\omega$ decays we get
\footnote{Notice that due to the method used to extract the two B.R.s, a
  significant correlation is present between the two values (see discussion in
  ~\cite{omegapi0})}:
\begin{equation}
B.R.(\omega\rightarrow\pi^+\pi^-\pi^0)=(90.24\pm0.19)\%
\end{equation}
\begin{equation}
B.R.(\omega\rightarrow\pi^0\gamma)=(8.09\pm0.14)\%
\end{equation}
significantly shifted with respect to current PDG values (respectively
(89.1$\pm$0.7)\% and $(8.90^{+0.27}_{-0.23})$\%).

\subsection{Update of hadronic cross-section measurement}
\begin{figure*}[t]
\centering
\includegraphics[width=65mm]{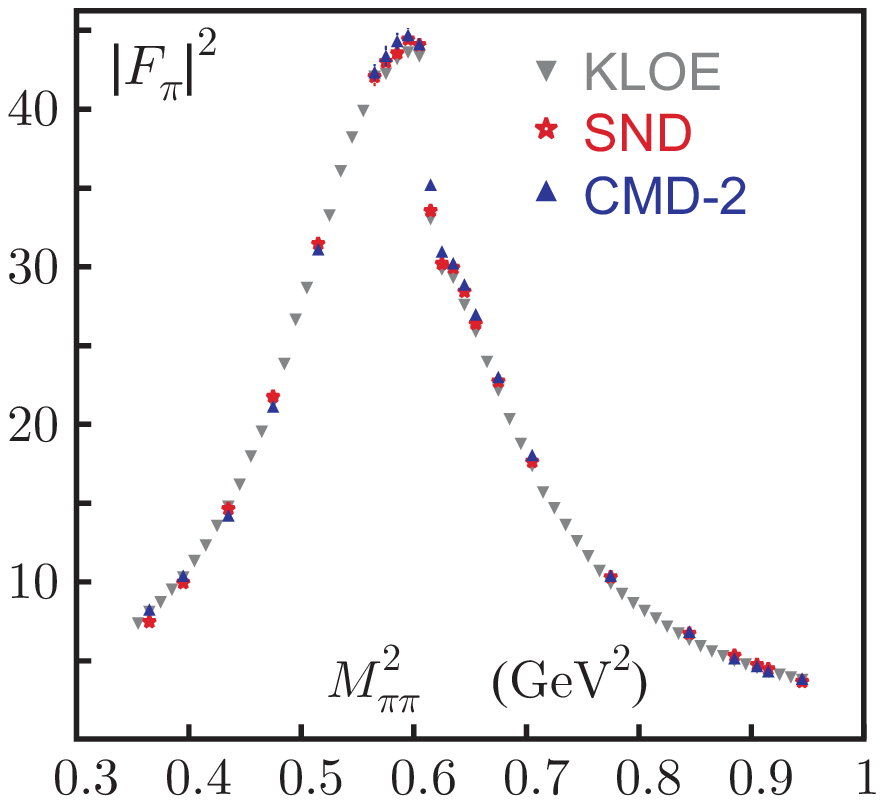}
\includegraphics[width=65mm]{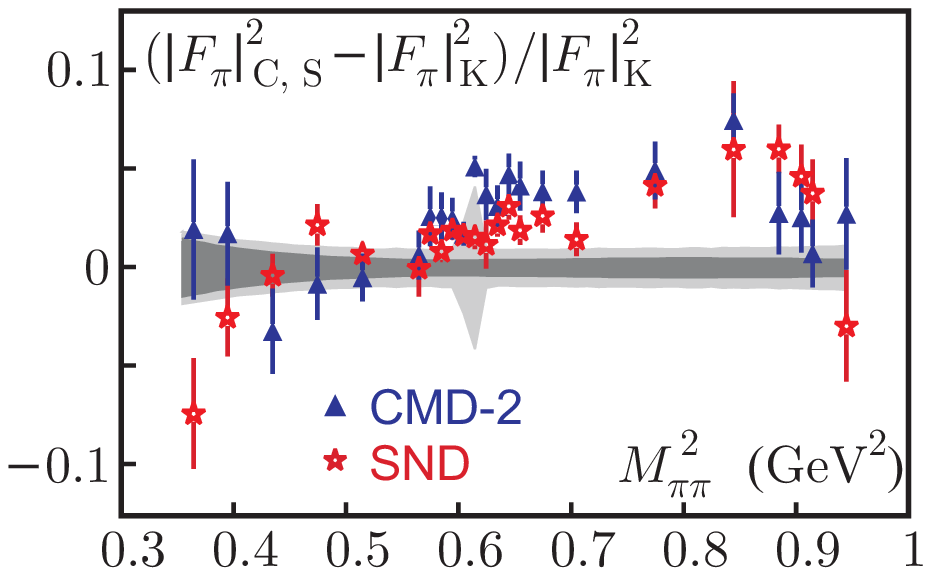}
\caption{(left) Square of the pion form factor as a function of
  $M_{\pi\pi}^2=q^2$ measured by KLOE (downward
  triangles) and by CMD-2 and SND. (right) Relative difference between KLOE
  and the two Novosibirsk experiments.} \label{hadronic}
\end{figure*}

The $e^+e^-\rightarrow\pi^+\pi^-$ cross-section for centre of mass energies
between the $\pi\pi$ threshold and 1 GeV is the main ingredient for the
theoretical evaluation of the hadronic contribution to the muon $g-2$~\cite{deRafael}. KLOE
has measured this cross-section in the $q^2$ region 0.35 $< q^2 <$ 0.95
GeV$^2$ ($q^2=s$)
using the ISR method~\cite{Kuhen}. A first result has been published 
based on a data sample of about 100 pb$^{-1}$~\cite{ppgold}. Here we present the result
of a new analysis on an independent data sample twice in statistics with
reduced systematic and theoretical errors. 
Fig.\ref{hadronic} shows the new KLOE results for the pion form factor compared to
the results obtained by the Novosibirsk experiments CMD-2 and SND at
VEPP-2M~\cite{Novosibirsk} that use the energy scan. From the measured
cross-section we obtain for the muon magnetic anomaly $a_{\mu}=(g-2)/2$:
\begin{equation}
a_{\mu}(0.35 < q^2 <
0.95)=(388.2\pm0.6_{stat}\pm3.3_{syst}\pm2.0_{th})\times 10^{-10}
\end{equation}
in agreement with the previous measured value and with the CMD-2 and SND values in
the same $q^2$ region. This result confirms the 3 $\sigma$ discrepancy between
the experimental value of $g-2$ and the theoretical
expectation~\cite{deRafael}. 

\section{OUTLOOK}
DAFNE is testing now a new machine scheme to increase luminosity. The first
results are encouraging and
KLOE will start a new run in summer 2009. The KLOE2 physics program
is based on an 
integrated luminosity exceeding 20 fb$^{-1}$ and on some substantial upgrades of
the detector. Among the physics items, other than several items in 
kaon physics, we mention $\gamma\gamma$ physics, improved
measurements in the $\eta$-$\eta'$ sector and observation of
$K\overline{K}\gamma$ final states.

\end{document}